\documentclass[review]{elsarticle}

\usepackage[breaklinks]{hyperref}
\usepackage{graphicx}
\usepackage{subfig}
\usepackage{epstopdf}
\usepackage{multirow}
\usepackage{array}
\newcolumntype{C}[1]{>{\centering\let\newline\\\arraybackslash\hspace{0pt}}m{#1}}
\usepackage{xcolor}
\usepackage{framed,multirow}

\usepackage{amssymb}
\usepackage{latexsym}

\usepackage{xcolor}
\definecolor{newcolor}{rgb}{.8,.349,.1}

\usepackage{natbib}
\usepackage{array, multirow, bigdelim, makecell, booktabs}

\usepackage{ulem}

\journal{Journal of \LaTeX\ Templates}

\biboptions{authoryear}

\begin{document}

\begin{frontmatter}
\title{Background model of Phoswich X-ray detector on board small balloon}

\author{Abhijit Roy}
\author{Ritabrata Sarkar\corref{cor1}}
\ead{ritabrata.s@gmail.com}
\ead{aviatphysics@gmail.com}
\author{Sandip K. Chakrabarti}
\ead{sandipchakrabarti9@gmail.com}

\address{Indian Centre for Space Physics, 43 Chalantika, Garia Station Road,
Kolkata 700084, W.B., India}
\cortext[cor1]{Corresponding author:
}

\begin{abstract}
We performed a detailed modelling of the background counts observed in a 
phoswich scintillator X-ray detector at balloon altitude, used for astronomical
observations, on board small scientific balloon. We used Monte Carlo simulation
technique in Geant4 simulation environment, to estimate the detector background
from various plausible sources. High energy particles and radiation generated
from the interaction of Galactic Cosmic Rays with the atmospheric nuclei is a
major source of background counts (under normal solar condition) for such
detectors. However, cosmogenic or induced radioactivity in the detector
materials due to the interaction of high energy particles and natural radioactive
contamination present in the detector can also contribute substantially to the
detector background. We considered detailed 3D modelling of the earth's
atmosphere and magnetosphere to calculate the radiation environment at the
balloon altitude and deployed a proper mass model of the detector to calculate
the background counts in it. The calculation satisfactorily explains the
observed background in the detector at 30 km altitude (atmospheric depth: 11.5
$g/cm^{2}$) during the balloon flight experiment from a location near
14.5$^{\circ}$N geomagnetic latitude.

\end{abstract}


\end{frontmatter}


\section{Introduction}
\label{sec1}

\label{sec:intro}
Astrophysical observations in X-ray or $\gamma$-ray energy band are carried 
out via space-borne or balloon-borne platforms to avoid the absorption of 
these rays in the earth's atmosphere. Proper understanding of the background
in X-ray or $\gamma$-ray detectors used in this purpose is very important; 
since the detectors work in a condition of low signal-to-noise ratio, due to 
low intensity radiation from the distant astronomical sources as received at
Earth. This knowledge is particularly necessary to reduce the background counts
in the detector by optimizing its design and also to extract the source signal
effectively from the observed data. The source of radiation background of
space-borne or balloon-borne detectors are physically the same, but the local
radiation intensity and spectra vary due to various space-weather phenomena
such as: radiation belts, South Atlantic Anomaly, solar energetic particles, 
cosmic radiation modified by the heliospheric electromagnetic-field varying 
with solar activity, atmospheric residual and albedo radiation etc. The
detector background strongly depends on the local space radiation environment,
which changes with the position and time of observation. For example, at
balloon altitude, the radiation belt and the South Atlantic Anomaly will have
nominal effects, while the atmospheric residual and albedo radiations will 
give a substantial contribution to the detector background.

The near-earth space radiation mainly composed of Galactic Cosmic Ray (GCR)
and Solar Particles Events (SPEs). The GCR consists of highly ionizing proton
($\sim$89\%), alpha ($\sim$10\%), small fraction of heavier nuclei and
electrons, which continuously bombard on the earth's atmosphere. The primary 
GCR at the top of our atmosphere undergoes modulation due to heliospheric
electromagnetic-wind and geomagnetic-field effect before entering the earth's
atmosphere. The GCR particles then interact with the atmospheric nuclei to
generate a cascade of secondary particles and radiation, which in turn creates 
a typical radiation intensity profile in the atmosphere. On the other hand, 
SPEs are originated from the sun via coronal mass ejection and/or solar 
flares, which can last from a few hours to a few days. The particles from 
SPEs are usually less energetic than primary GCR and may extend up to a few 
GeVs during strong solar activity.

The background count of an X-ray detector used in astronomical observations,
on board a balloon-borne platform, suffers most from these varying atmospheric
secondary radiations. The total background in the detector can be considered to
be due to integrated contributions from: (i) direct energy depositions from the
aforementioned energetic particles and radiation in the active detector; (ii)
locally produced secondary particles and radiation generated via interaction
with the passive materials present in the vicinity of the active detector; (iii)
radioactive decay of the cosmogenic radio-nuclides produced in the detector due
to spallation and activation of the nuclei of the detector materials by the
highly energetic Cosmic Ray (CR) particles; (iv) radiation from the decay of
natural radio-isotopes present in the detector and its surrounding materials 
\citep{dean91}. It is worth mentioning another contribution to the
balloon-borne detector background from the terrestrial radioactivity. But this
component is prevalent only up to 2-3 km ($\sim$750-670 $g/cm^{2}$) from the
Earth's surface (see Fig. \ref{fig:detLCwithAlt}a) and diminish rapidly due to
atmospheric absorption.

The Monte Carlo (MC) simulation technique is an efficient method to accurately
model the detector background and has been extensively used for this purpose.
However, simulation of the radio-activated background component is quite
complicated task which has been successfully implemented, for example, by
\cite{weid05} and \cite{odak18}, to explain the observed background of
satellite-borne X-ray detectors. For heavily shielded, ground based, low
background, dark matter probing experiments the observed counts have also been
explained by considering the radio-activation along with the natural radioactive
background of the detector and its surrounding material by several authors
\citep{adhi17, adhi18, abde19}.

Here, in this work, we use a similar kind of methodology to explain the observed
background of a balloon-borne X-ray detector using extensive MC simulation with 
Geant4 simulation toolkit \citep{GEANT4}.

The scientific ballooning team in Indian Centre for Space Physics, Kolkata, 
India, has been involved in small balloon-borne experiments to observe the 
extraterrestrial radiation for the past several years \citep[and references 
therein]{chak17}. Because of the imposed payload weight constraint, the 
detectors used in these experiments can neither use heavy shielding nor any
pointing devices. This affects the observational results in terms of poor
reduction of the physical background. This constraint also does not permits us
for simultaneous measurement of background during the observations of
astronomical radiation sources. Furthermore, the detector undergoes diverse
radiation environment during its operation as it ascends gradually towards the
burst altitude of the balloon and while coming down thereafter. So, an accurate
background modeling is a necessity for extracting any meaningful results out of
our observed data. We have been mostly relying on the empirical solutions for
the background estimation of the detectors \citep{sark19, sark20}. Here in this
work, we successfully explain the total observed detector background by
considering the sources of all plausible major background components through
extensive MC simulation.

In Sec. \ref{sec:expt}, first we briefly describe the experimental observation
considered in this work to study the background counts. Then we discuss the
simulation methodology to model the detector background in Sec. \ref{sec:simu}.
We show the results of the background modelling and its comparison with the
observed counts in Sec. \ref{sec:results} and conclude the results in Sec.
\ref{sec:conc}.

\section{Experimental Overview}
\label{sec:expt}
We have been using small and light-weight X-ray detectors on board small sized
($\sim$4000 cubic meters) plastic-balloon platform for the observation of
astronomical sources in X-ray energy band. The balloon carries the payload of
about 5 kg weight, up to $\sim$40--42 km ($\sim$2.9--2.2 $g/cm^{2}$) above the 
earth's surface where the residual atmosphere is small enough to allow incoming 
X-rays from the extraterrestrial sources. Usually, there is no cruising level 
for this kind of balloon flights as there are no ballast and valve system used 
in these carriers. The balloon goes up, get ruptured at certain 
level due to excess internal pressure and gradually comes down with the 
payload with an average speed of about 4 m/s (may vary drastically, 
particularly during descent, depending on the nature of rapture of the balloon) 
\citep{chak17}. The main X-ray detection unit, used in the experiment under 
consideration, is a scintillator detector consisting two crystals (3 mm thick 
NaI(Tl) and 25 mm thick CsI(Na), both of 116 mm diameter) in phoswich 
combination and optically coupled with a Photo-Multiplier Tube (PMT). We use a 
collimator, made of 0.5 mm thick tantalum providing 15$^{\circ}$ Field-of-View 
(FoV), to minimize the off axis radiation. In this study, we 
consider only the events with its full energy deposited in the primary (NaI) 
crystal to discard the background counts due to partial energy depositions. 
This has been achieved by considering the anticoincidence technique using the 
pulse shape information of the events recorded by the detector readout system. 
The pulse shape information dictates whether the signal is from the NaI crystal 
or have energy depositions in both the crystals (or may be in CsI crystal 
alone).

The background count of the detector provides a minimum detection limit above
which any astronomical source could be detected. The sensitivity calculation of
the detector shows that the residual atmosphere allows the detector to observe
typical astronomical X-ray sources (like Crab pulsar) above the altitude of 30
km (11.5 $g/cm^{2}$) \citep{sark20}. Here, in this work, we use the results of
the mission {\it Dignity 113} which was launched from Muluk, West Bengal, India
(latitude: 23.64$^{\circ}$N; longitude: 87.71$^{\circ}$E; geomagnetic latitude:
14.5$^{\circ}$N) on November 23, 2019 at 05:41 UT. Further information of the
experiment could be found in Table \ref{Tab:InfoMis}. The operating energy 
range of the detector is $\sim$10--100 keV. However, for the present analysis, 
we have considered only $\sim$10--80 keV range to avoid the computational 
complications due to the collimator transparency at higher energy range.

\begin{table*}
\caption{Experimental overview}
\label{Tab:InfoMis}
\centering
\begin{tabular*}{\textwidth}{@{\extracolsep{\fill}}ll@{}}

\hline\noalign{\smallskip}
Main detector 				& phoswich detector having 3 mm NaI crystal and \\
					& 25 mm CsI crystal\\
Collimator 				& 0.5 mm thick Ta collimator with 15$^{\circ}$ FoV\\
Other ancillary instruments 		& GPS system, payload tracker system, attitude \\
					& measurement system\\
Total payload weight 			& 5.610 kg\\
Detector orientation 			& towards local zenith\\
Payload carrier				& one meteorological plastic balloon (length: $\sim$ 25 m, \\
					& weight: 8.075 kg)\\
Date of launch 				& Nov 23, 2019\\
Time duration 				& 05:41 -- 13:33 UT\\
Launching location 			& Muluk, WB, India (Lat: 23.64$^{\circ}$N, Lon: 87.71$^{\circ}$E)\\
Maximum height attains 			& 40.09 km (at 07:50 UT)\\
Landing location 			& Bali Number 2, WB, India (Lat: 22.81$^{\circ}$ N, Lon: \\
					& 87.77$^{\circ}$ E)\\
\noalign{\smallskip}\hline
\end{tabular*}
\end{table*}

\subsection{Detector calibration and resolution}
\label{ssec:cal}
\begin{figure*}
\centering
\resizebox{0.45\textwidth}{!}{
\includegraphics[width=.45\textwidth]{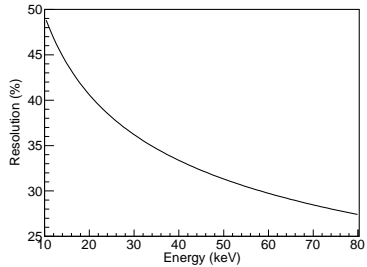}
}
\caption{Energy resolution of the detector as a function of energy.}
\label{fig:engResEffArea}
\end{figure*}

During the mission preparation phase, the detector is thoroughly tested under
different experimental conditions (temperature/pressure) in the laboratory. The
energy calibration of the detector is done using standard calibration radiation
sources ($^{133}$Ba, $^{137}$Cs, $^{152}$Eu and $^{241}$Am) to find the
energy-channel relation. The energy resolution of the detector is obtained by
Gaussian fitting of the decay peaks of the calibration sources. (The energy 
resolution being defined as full-width at half-maximum divided by the mean 
energy of the Gaussian peaks.) The energy dependence of the resolution was 
found to vary as $1.944((E+0.226)/0.083)^{-0.285}$ (E being the energy 
in keV) and is graphically shown in Fig. \ref{fig:engResEffArea}. 
The detailed procedure and results of the extensive laboratory testing can be 
found in \cite{bhow19}. The detector resolution is necessary for the spectral 
analysis of the detector data, as well as to compare the simulated and observed 
results.

\subsection{Observational data}
\label{ssec:obs}
\begin{figure*}
\centering
\resizebox{1.0\textwidth}{!}{
\subfloat[]{\includegraphics[width=.45\textwidth]{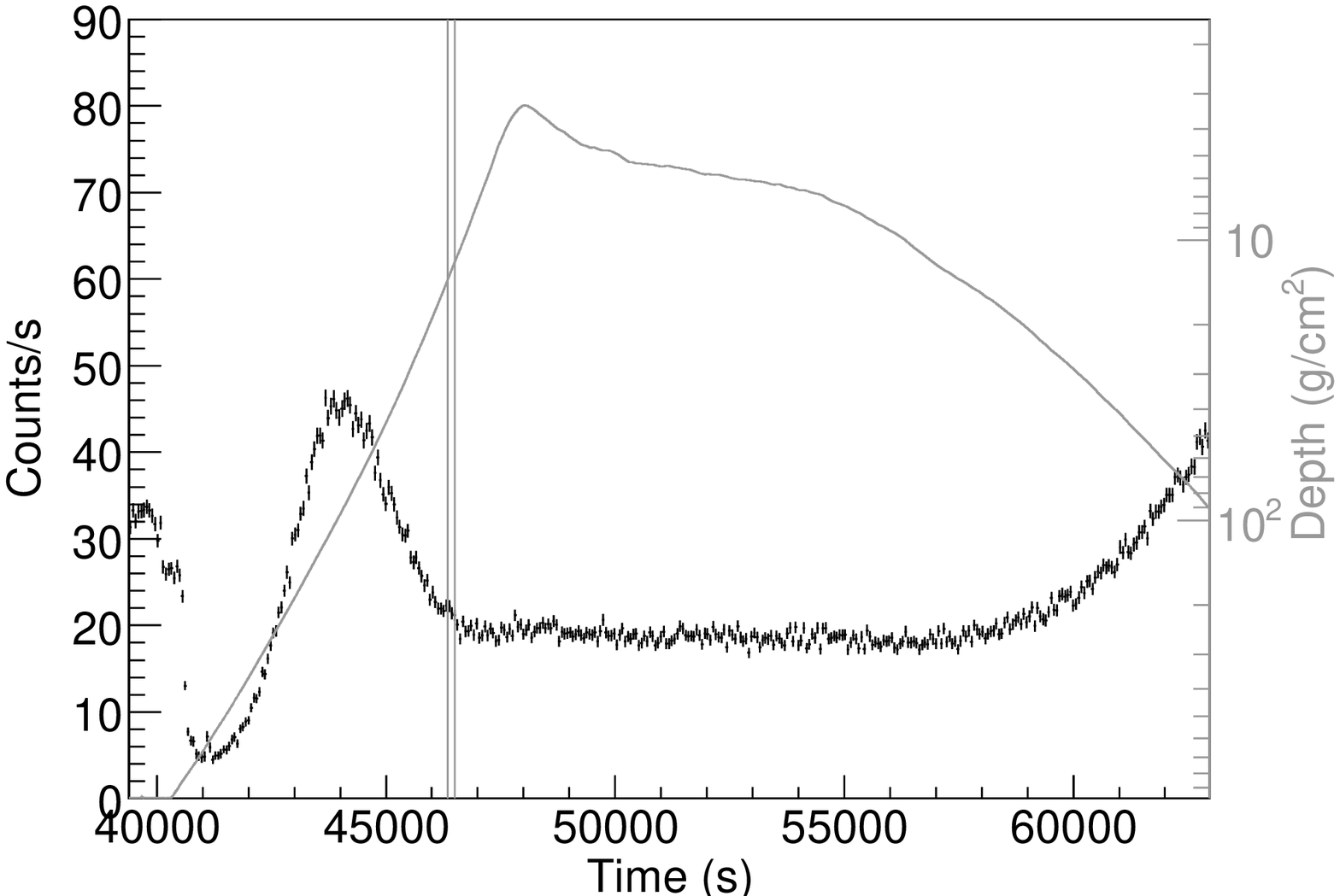}}
\subfloat[]{\includegraphics[width=.45\textwidth]{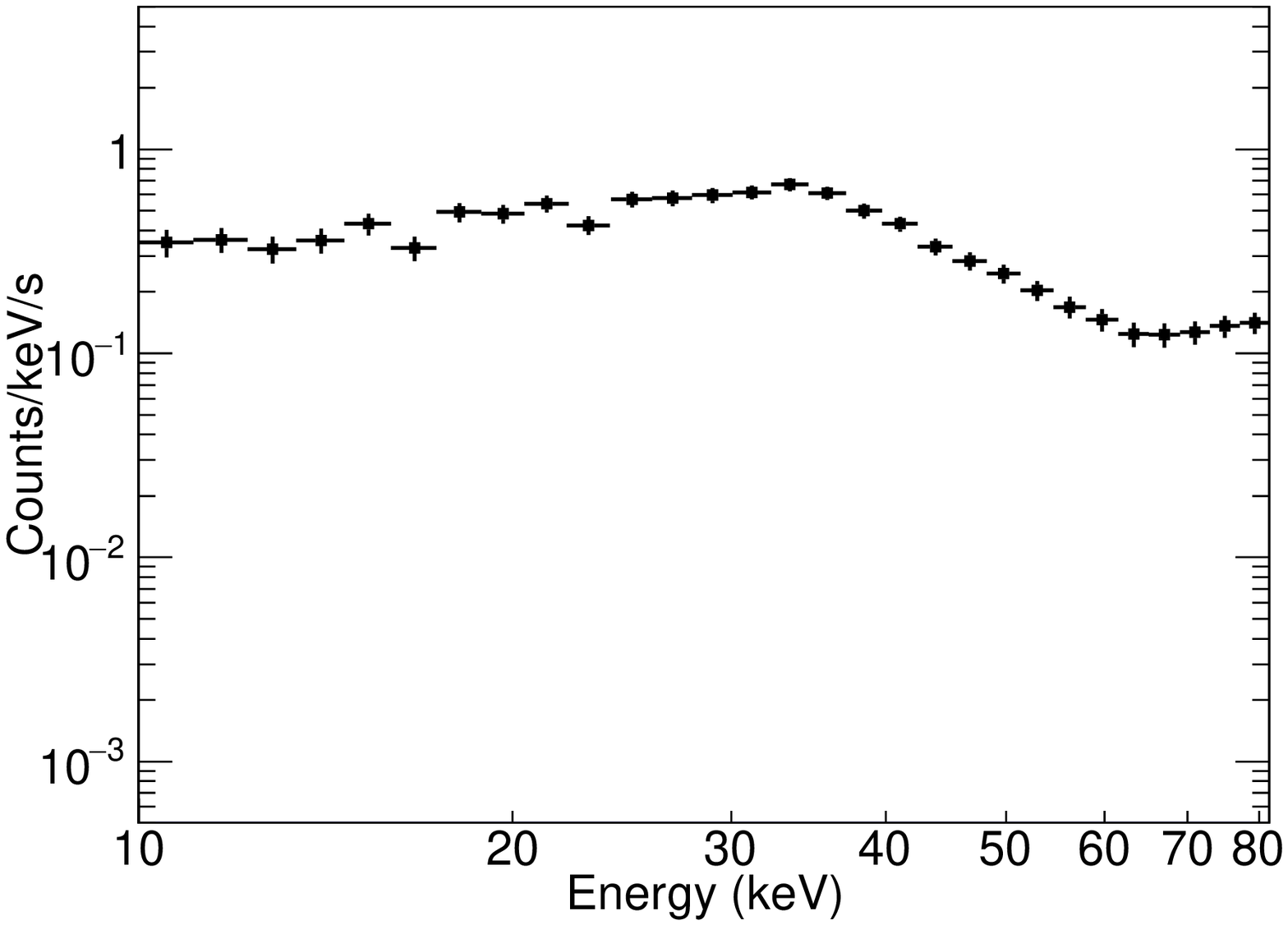}}
}
\caption{(a) Integrated atmospheric radiation count rate in the 10-80 keV
energy range detected by the phoswich detector during the entire mission flight
(black points), along with the payload altitude profile (gray line). Vertical
lines showing the region around 30 km from where the detector data is considered
for background analysis. (b) The background spectrum of the detector at 30 km
altitude.}
\label{fig:detLCwithAlt}
\end{figure*}

The detector data is recorded and stored, on board in a memory-card provided
to the payload, throughout the mission life from launching to landing. The
observed radiation count rate, averaged over 1 minute time bin, is shown in 
Fig. \ref{fig:detLCwithAlt}a, along with the instantaneous payload altitude
starting from the balloon launch. The peak in the atmospheric radiation count
at about 16 km ($\sim$104 $g/cm^{2}$) is due to the Regener-Pfotzer maximum and 
has been discussed in more details in \cite{sark17}. Due to some unresolved
instrumental problem, the detector ceased to operate at around 17 km ($\sim$88
$g/cm^{2}$) altitude during the payload descent. However, the data obtained 
from the rest of the mission is adequate for the purpose of the current work. 
We consider the radiation spectrum due to atmospheric background counts at 
30 km ($\sim$11.5 $g/cm^{2}$) altitude (marked by vertical lines in Fig.
\ref{fig:detLCwithAlt}a). Figure \ref{fig:detLCwithAlt}b shows the detector
spectrum (total count rate) due to the overall radiation at 30 km.

\section{Simulation of Detector Background}
\label{sec:simu}

\subsection{Simulation methodology}
\label{ssec:simumethod}
For the simulation of background counts in a detector on board a balloon-borne
platform, we mainly need to consider the following points: the radiation
environment surrounding the detector, a proper mass model of the detector and
its surrounding materials and the physics processes which dictates the
interactions of the radiation or particles with the detector. To take account 
of the total observed background counts in the detector, we need to simulate 
background counts due to: (i) instantaneous energy depositions in the detector 
from the high energy particles and radiation (along with those produced in the 
materials close by the detector); (ii) the prompt and delayed radioactive decay 
of cosmogenic induced radionuclides produced in the detector due to activation 
and spallation of nuclei in the detector materials by the 
high-energy particles and (iii) decay of long-lived natural radioactive 
isotopes present in the detector materials \citep{pete75}. 

First of all, we simulate the background counts from direct energy depositions
of atmospheric secondary CR particles in the detector considering: a proper
mass model of the detector, radiation environment and physical interaction
model. Calculation of the radiation environment model of the detector, at the
position of interest, is done by considering detailed atmospheric and
magnetospheric models. We considered the NRLMSISE-00 standard atmospheric model
\citep{pico02} to describe the earth's atmosphere up to 100 km from the earth's
surface with proper input parameters for location, time and solar condition. 
The inner magnetic field at the vicinity of the earth's surface due to earth's
magnetism is calculated using the 12th generation IGRF model \citep{igrf15}.
The external magnetic field which depends on the interplanetary magnetic field
and solar condition is calculated using Tsyganenko model \citep{tsyg16}. The
detailed procedure of this simulation can be found in \cite{sark20}. 
Considering the location and time (at quiet solar condition) of the experiment,
we use only the GCR spectrum with proper model \citep{vos15,herb17}, as the
primary input in the simulation without any SPE contribution. The spectral and
directional distribution of the produced secondary particles and radiation at
the relevant atmospheric location are used for further calculation of the
background counts through simulation of radiation interaction in the detector.

To account the background due to cosmogenic radio-activation, we calculate the
production of radio-nuclides in the detector via activation or spallation
process, for each simulation event (i.e., incident particle). The production of
cosmogenic radioactive isotopes, after the irradiation of detector mass model
are calculated using the {\it G4RadioactiveDecay} process provided in the 
Geant4 toolkit. Whenever any cosmogenic radionuclide is produced, we record the
information of their mass number (A), atomic number (Z), excitation energy
(ExEng), kinetic energy, position, momentum and pdg encoding 
(MC particle numbering scheme according to \cite{pdg20}). 
Furthermore, to avoid the double counting in the energy deposition, we kill 
the track of that radionuclide and all its subsequent tracks. After recording 
all the cosmogenic radio-nuclides from the irradiation of all the GCR 
interactions, we separately simulate these induced radioactive nuclei allowing 
them to decay in proper decay channel(s), considering their position, momentum 
and excitation energy in the detector materials. One caveat is 
worth mentioning here, if the cosmogenic radionuclides (parent) induced in
the active crystal by the GCR, decays within the characteristic signal decay 
time of the crystal itself, the whole energy deposition from the direct 
interaction and radioactive decay will provide a single event signal. Although, 
in principle, this effect should be taken into consideration during 
digitization of the simulated event, in the current procedure we have not 
considered this effect. Practically, in the current procedure this does not 
affect the result substantially, since in this analysis we found no significant 
abundance of induced parent radio-isotopes which promptly decays to give rise 
this pile-up effect. This can be seen from the half lives of the enlisted 
parent isotopes in Table \ref{tab:cosmoTab} which give the major contributions 
in our energy range of interest. However, the pile-up effect during energy 
depositions from different decay stages of the subsequent decay of induced 
parent isotopes has been considered inherently in the simulation, if the 
tracks have energy depositions in the active crystal within its signal decay 
time.

The internal background component of the detector due to natural radioactive 
isotopes in the detector material is calculated by generating the 
radio-nuclides randomly inside and on the surface of the primary crystal (NaI) 
of the detector and allow them to decay according to their proper decay 
channel(s), following the same approach as mentioned by \cite{adhi17, adhi18}.

\subsubsection{Particle generation model}
\label{sssec:partgen}
\begin{figure*}
\centering
\resizebox{1.0\textwidth}{!}{
\subfloat[Downward]{\includegraphics[width=0.45\textwidth]{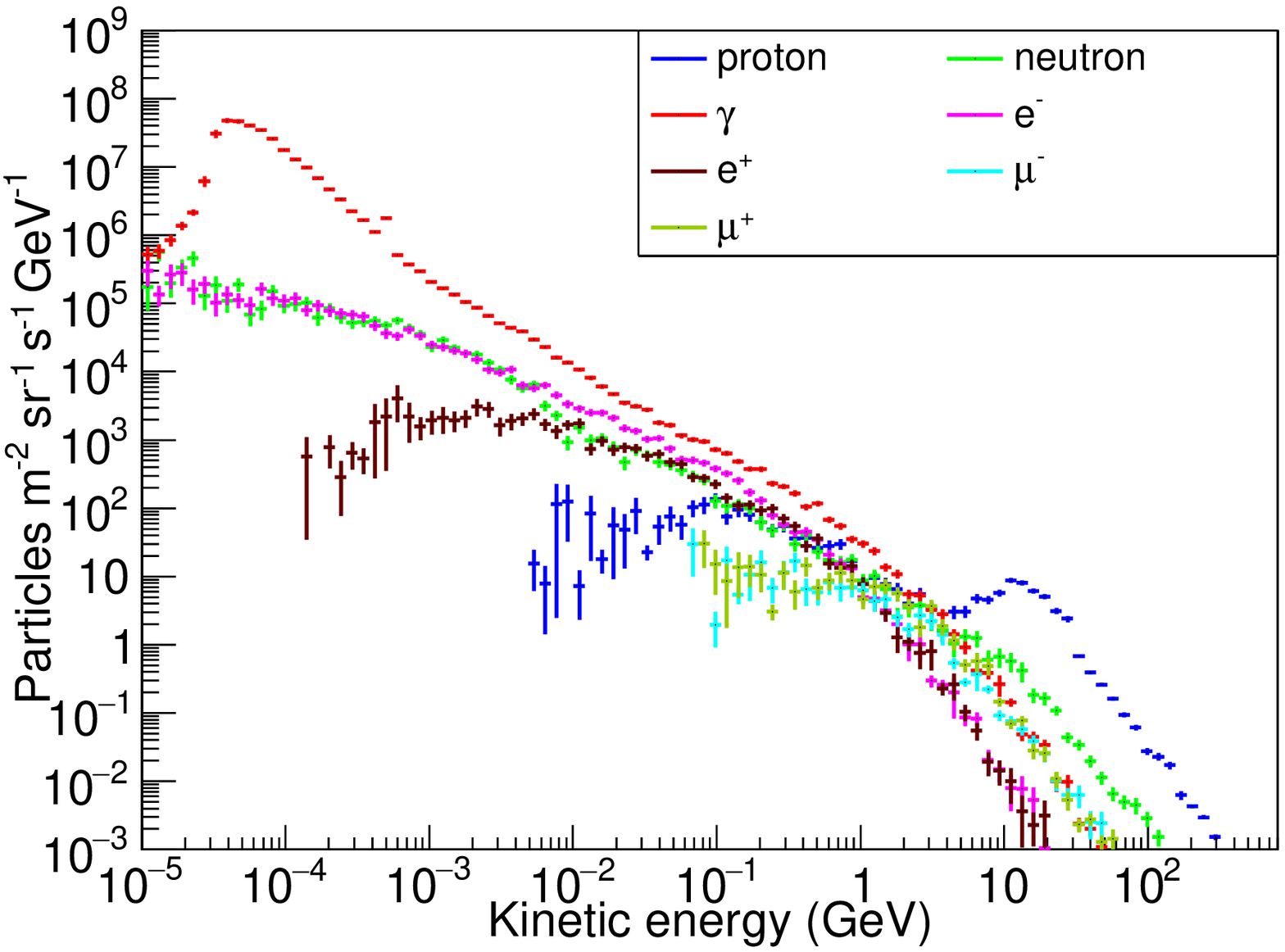}}
\subfloat[Upward]{\includegraphics[width=0.45\textwidth]{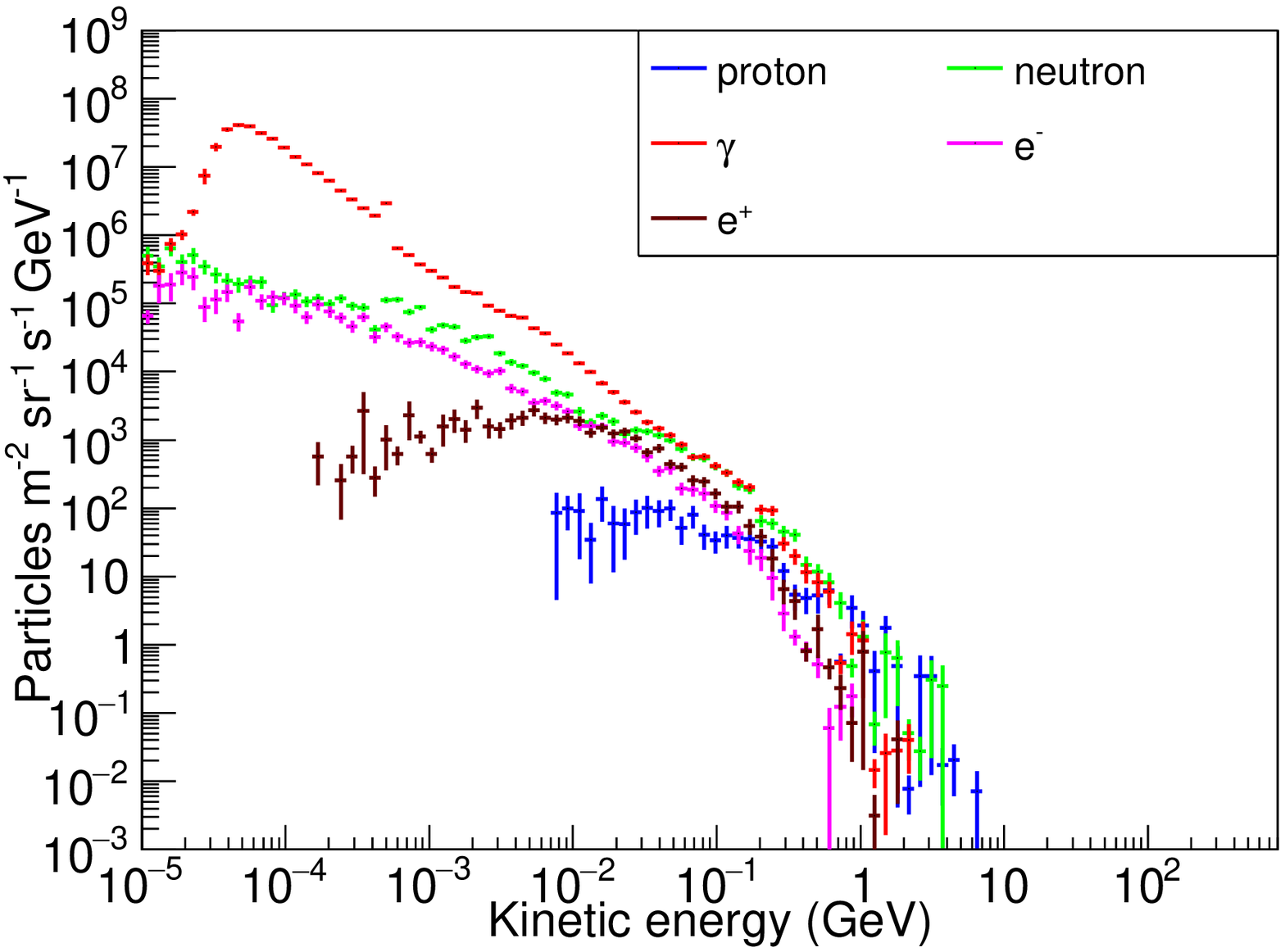}}
}
\caption{Flux distribution of different secondary CR (and 
residual primary) particles at 30 km altitude in the geomagnetic latitude 
range of 11.5$^\circ$ -- 17.2$^\circ$, for both downward and upward particles 
with respect to local zenith.}
\label{fig:seconEngFlux}
\end{figure*}

During the ascent and descent of the balloon in the atmosphere, the detector
undergo irradiation from diverse radiation environment as can be seen from
the radiation intensity profile in Fig. \ref{fig:detLCwithAlt}a. So, an 
accurate modelling of radiation environment due to the interaction of CR
particles and radiation with the atmosphere at the payload altitude under
interest is necessary. For this purpose, the flux distribution of major
atmospheric secondary GCR particles along with the residual primaries at that 
altitude are calculated according to the simulation procedure described in 
\cite{sark20}. The simulation procedure considers all the updated atmospheric, 
geomagnetic field models and primary GCR flux model to propagate and transport 
the GCR and cosmic diffuse gamma-ray background photons \citep{ajel08, weid99} 
into the atmosphere, considering proper modulation of the charged particles by 
the magnetosphere. The calculated flux distribution of major species of 
residual primary and secondary CR particles generated at 30 km altitude in the 
earth's atmosphere is shown in Fig. \ref{fig:seconEngFlux} separately for 
downward and upward particles. The directional distribution 
of the particles were calculated as a function of the cosine of zenith angle
($\cos\theta_Z$) and were fitted using (2nd or 3rd degree) polynomial functions 
of $\cos\theta_Z$ for different particles. For the simulation of radiation 
interaction in the detector considered in this work, we irradiated the detector 
with 10$^6$ particles of each species for both downward and upward particles 
(from two hemispherical surfaces), according to the particle 
flux distributions given in Fig. \ref{fig:seconEngFlux}, with kinetic energy 
ranging from 10 keV to 800 GeV. We conserved the directional distribution of 
both the downward and upward particles obtained from the previous simulation 
of secondary particle generation, during the irradiation of the detector. 
The directional distribution is achieved by biasing the random 
distribution of the source positions on the two hemispheres according to the 
$\cos\theta_Z$ dependent polynomial functions from the previous step of the 
simulation.The direction of the particles generated at each points on the 
source surface are also random in cosine distribution w.r.t. the surface 
normal at that position. Considering the geometry of the source distribution 
and position of the detector, we resitricted the angular random distribution 
in 15 deg from the surface normal direction at the position of generation, 
to increase the probability of the tracks hitting the detector.

\subsubsection{Detector mass model}
\label{sssec:detmodel}
\begin{figure}[!ht]
\centering
\resizebox{0.45\textwidth}{!}{
\includegraphics{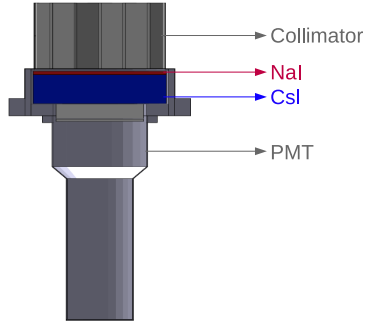}
}
\caption{Cut-away view in YZ-plane of the phoswich detector module considered
in the simulation.}
\label{fig:pswDet}
\end{figure}

Modelling the proper mass distribution of the detector and its surrounding
materials is very important, as the interaction of the high energy particles and
radiation depends on this material distribution which, in turn, produce other
secondary particles and induced radio-nuclides contributing to the background
counts. The payload box used in the experiment is made by Styrofoam and there is
no significant amount of high density materials used around the detector, apart
from the thin aluminum frame to hold various payload components and PCBs for the
electronic circuits \citep{bhow19}. So, here in this simulation we only consider
the mass model of the detector (with PMT) along with the collimator and
shielding materials. The mathematical mass model of the phoswich detector is
constructed in the Geant4 simulation environment, using the realistic dimension
and composition of the detector. The cut-away view in YZ-plane of the detector
mass model used in the simulation can be seen from Fig. \ref{fig:pswDet}.

\subsubsection{Physical interaction model}
\label{sssec:phymod}
Considering the requirements of the current simulation, we use {\it
QGSP\_BIC\_AllHP} with updated {\it TALYS-based Evaluated Nuclear Data Library
(TENDL)} among the many physics lists available in Geant4. This physics list is
capable of handling the hadronic interactions and production of radio-nuclides
which is very important for this kind of simulation. {\it QGSP\_BIC\_AllHP} is
identical to {\it QGSP-BIC-HP} physics list except for the low energy charged
particles interactions \citep{g4pl20}. For electromagnetic interactions,
{\it electromagnetic\_options3} is used which is capable of handling the
interactions well in the range of detection limits of the current experiment.
The {\it QGSP\_BIC\_AllHP} physics list can handle hadronic interactions up to
the energy of 10 TeV and the electromagnetic physics list covers the energy
range from 0 to 100 TeV which is well within our required simulation range. For
proper decay of radionuclide within the Geant4 toolkit, the {\it
RadioactiveDecay5.3} and {\it PhotonEvaporation5.3} data sets are used. We
considered the production cut value of 1 $\mu$m for gamma, electron, positron
and proton during the secondary track generation in the simulation.
 
\subsection{Signal digitization and flux normalization}
\label{ssec:norm}
NaI is the primary crystal of the scintillator detector used here for the
radiation count. Any considerable energy deposition (low-level discriminator
voltage is set to give trigger threshold energy around 10 keV) gives rise to a
signal in the detector. However, a relatively thick CsI crystal is placed just
beneath the primary crystal to eliminate the Compton background and partial
energy depositions in the primary crystal. We implemented the anticoincidence
technique also in the signal digitization routine of the 
simulated energy depositions, just as is done during the analysis of the 
observed data.

In order to compare the simulated background with the observed background during
the experiment, we need to modulate the simulated energy deposition considering
the detector resolution during the signal digitization. The deposited energy 
inside the NaI crystal is modified according to the detector resolution shown 
in Fig. \ref{fig:engResEffArea}, to convert them into channel energy of the 
actual detector.

The normalization of the energy spectra in the detector from the direct energy 
depositions of CR particles is done as follows. The normalization factor is 
calculated by integrating the incident particle spectra over the energy range 
divided by the number of simulated events. Then, multiplied by the integrated
area of event generation surface (i.e., area of each hemisphere) and the solid 
angle integrated over the domain considered in the simulation (integration of 
azimuthal angle over $2\pi$ and polar angle up to 15$^{\circ}$ considering 
cosine distribution). Thus we get the unit same as the observed count rate of 
the detector.

The normalization factor of the detector counts is achieved by integration over 
the source surface (i.e., area of the hemisphere) and the solid angle domain
considered in the simulation (integration of phi over 2pi and theta up to 15 
deg considering cosine distribution).

The situation is different for the calculation of normalized 
background counts due to energy depositions from the cosmogenic induced 
radioactivity in the detector. Here, we modify the simulated energy deposition
from the decay of radioactive isotopes using the detector resolution as usual. 
But the normalization of the counts can not be done by simply considering the
incident particle spectra in the same way as was done for direct energy 
depositions. Due to the diverse radiation-environment history witnessed by the 
detector during its flight, it becomes very complicated to exactly calculate 
the normalization factor for the background counts from the induced radioactivity. 
So, we consider the following procedure to calculate the 
normalized contribution from the induced radioactivity. The effective 
background due to induced radioactivity depends on their production rate due 
to the radiation flux incident on the detector at each layer in the atmosphere 
and subsequent decay of these radionuclides. This can be a complicated process 
to calculate. Instead, we consider a simplified assumption of fixed incident 
radiation flux at the height of current calculation (30 km) shown in Fig. 
\ref{fig:seconEngFlux}, to produce the initial list of parent radioisotopes by 
irradiating the detector with these radiation flux. Then, these induced 
radioisotopes are decayed according to their respective decay channel(s) and 
half lives and deposited energy is calculated in the detector crystal to give 
background counts as mentioned in Sec. \ref{ssec:simumethod}. However, we use 
only those radionuclides which deposit the energy in our region of interest 
($\sim$10-80 keV). The deposited energy of each individual radioisotopes are 
then modified according to the detector resolution. Then, all these relevant 
energy deposition peaks in the energy range of interest are fitted with the 
observed data with independent normalizations (also considering other 
background contributions) to get the corresponding normalization factors of 
the individual peaks from different induced radio-isotopes. (See Fig. 
\ref{fig:CosmoComp} and \ref{fig:bkg} and discussions in Sec. 
\ref{sssec:inducedCont}).

In case of background counts from the long-lived natural radioactive isotopes 
in the detector crystal, the problem of normalization factor persists due to
the unknown fraction of radioactive contamination, which mainly 
occurs during the crystal growth. Also, because of the poor resolution of the 
detector, radiation lines from the characteristics decay peaks cannot be 
segregated and estimated which is possible with the cryogenic solid state 
detectors. So, we need to rely on the calculation of the normalization factor 
from the fitting of calculated flux (along with other 
background components) with the observed data.

\section{Results and Discussions}
\label{sec:results}

\subsection{Contributions from different background sources}
\label{ssec:contributions}
We now discuss the background contributions in the detector 
from different sources as mentioned earlier. The different energetic particles 
from the atmospheric radiation environment including the residual primary 
particles interact with the detector material distribution (considered 
in Sec. \ref{sssec:detmodel}) to deposit their energy directly in the detector 
crystal or via production of induced radionuclides. These induced radionuclides 
(along with natural radioactive isotopes present in the crystal) subsequently 
decay according to their allowed decay channel(s) with corresponding half lives 
and deposit energy in the crystal. Since the observation is done at low 
geomagnetic region where the rigidity cut-off is high and there is no 
considerable solar event at the time of observation, we neglect the 
contribution of solar cosmic-rays to the background counts.

\subsubsection{Contribution from direct energy deposition of secondary cosmic
rays}
\label{sssec:dirCont}
\begin{figure*}
\centering
\resizebox{1.0\textwidth}{!}{
\subfloat[]{\includegraphics[width=0.45\textwidth]{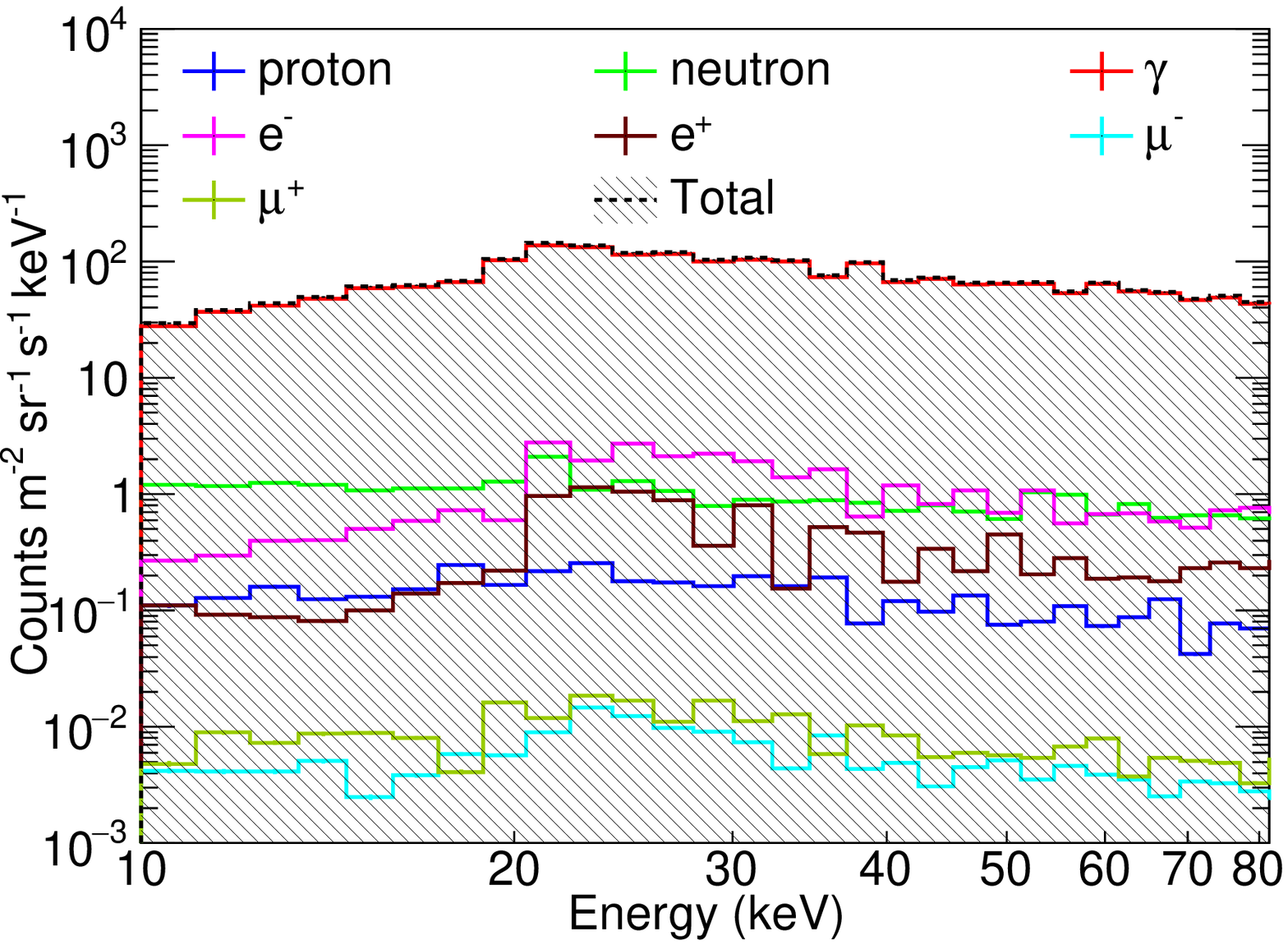}}
\subfloat[]{\includegraphics[width=0.45\textwidth]{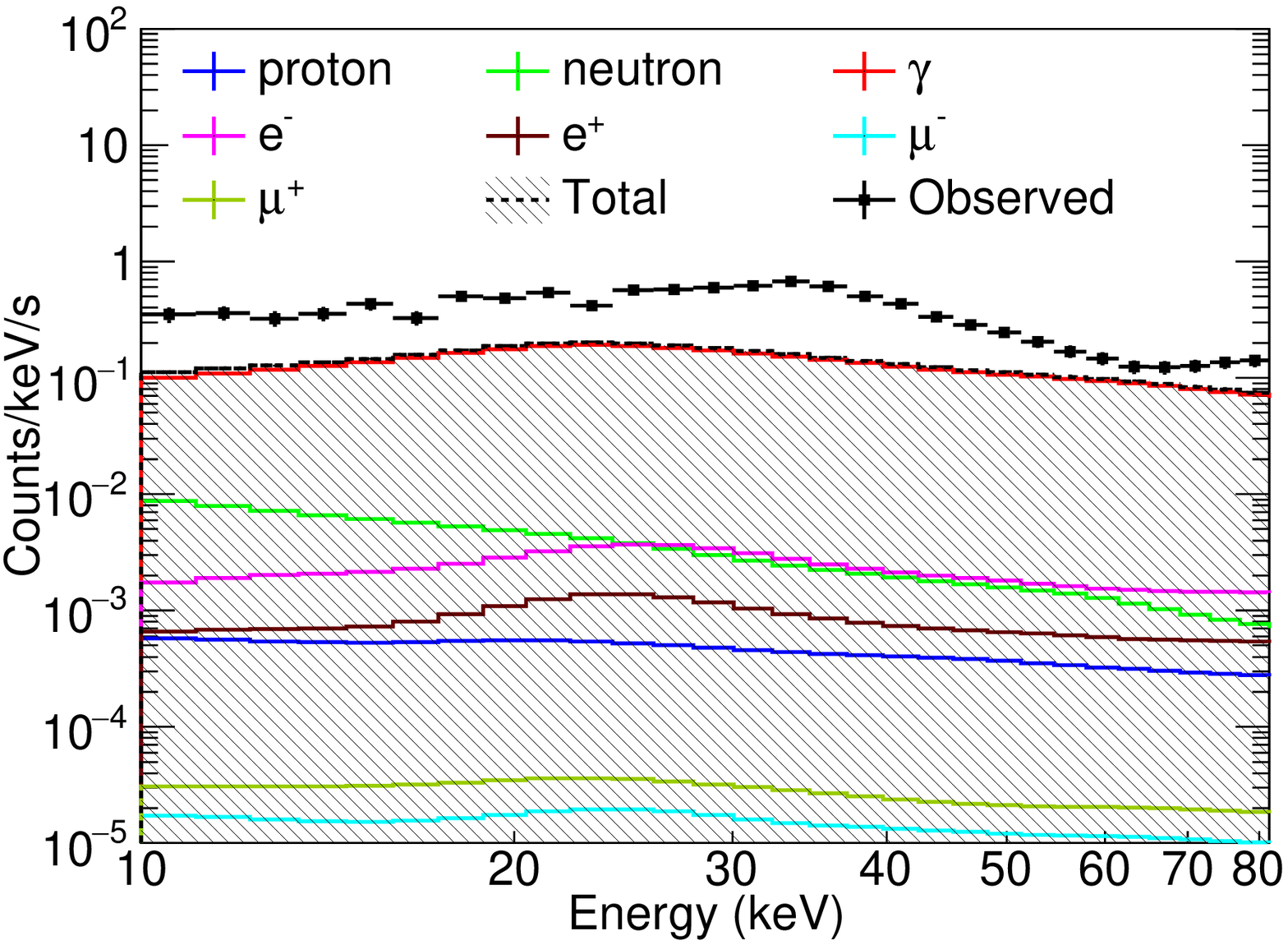}}
}
\caption{(a) Simulated particle flux in the detector due to different secondary
CR particles in NaI crystal, with considering anticoincidence technique. (b)
Contribution of different CR particles to the detector counts along with the
observed detector counts at 30 km altitude.}
\label{fig:GCRComp}
\end{figure*}

Simulated flux distribution in the detector due to the major components of
secondary CR particles in the atmosphere are shown in Fig. \ref{fig:GCRComp}a.
Whereas, in Fig. \ref{fig:GCRComp}b, we show the count-rate distribution in the
detector due to contributions from all the individual particle and radiation
species along with the total observed count rate in the detector. We also
tabulate the simulated count rates and corresponding relative contributions to
the total observed background in Table \ref{tab:GCRContTab}. From these
simulation results, it is quite evident that atmospheric $\gamma$-ray is the
principal contributor to the total observed detector background among all the
secondary GCRs, and it is quite justified because of the use of thin NaI crystal
and anticoincidence technique for background reduction. Due to 
their energy deposition nature, most of the other particles (than $\gamma$-rays 
and neutrons) with lower energies will be absorbed in the materials covering 
the detector crystals, while the higher energy particles are likely to deposit 
their partial energy both in NaI and CsI crystals, thus being removed from the 
background counts by anticoincidence requirement. From Fig. \ref{fig:GCRComp}b, 
it is also apparent that direct energy depositions from atmospheric CRs 
contribute only a part of the detector background and we need to include the 
background counts from other sources as well.

\begin{table}
\caption{Simulated count rates (Counts/s) from different atmospheric particles
and their relative contribution (Rel. Cont.) to the total observed count rate}
\label{tab:GCRContTab}
\centering
\begin{tabular*}{\columnwidth}{@{\extracolsep{\fill}}ccc@{}}

\hline\noalign{\smallskip}
Particle	& 	Counts/s  	&		Rel. Cont. (\%)   \\ 
\noalign{\smallskip}\hline\noalign{\smallskip}
p           &    0.028  &    0.126           \\
n           &    0.178  &    0.787           \\
$\gamma$    &    8.552  &    37.76           \\
$e^{-}$     &    0.151  &    0.666           \\
$e^{+}$     &    0.054  &    0.241           \\
$\mu^{-}$   &    0.000	 &   0.004           \\
$\mu^{+}$   &    0.001	 &   0.007           \\
\noalign{\smallskip}\hline
\end{tabular*}
\end{table}

\subsubsection{Contribution from induced cosmogenic radio-activation}
\label{sssec:inducedCont}
\begin{figure}[!ht]
\centering
\resizebox{\columnwidth}{!}{
\includegraphics{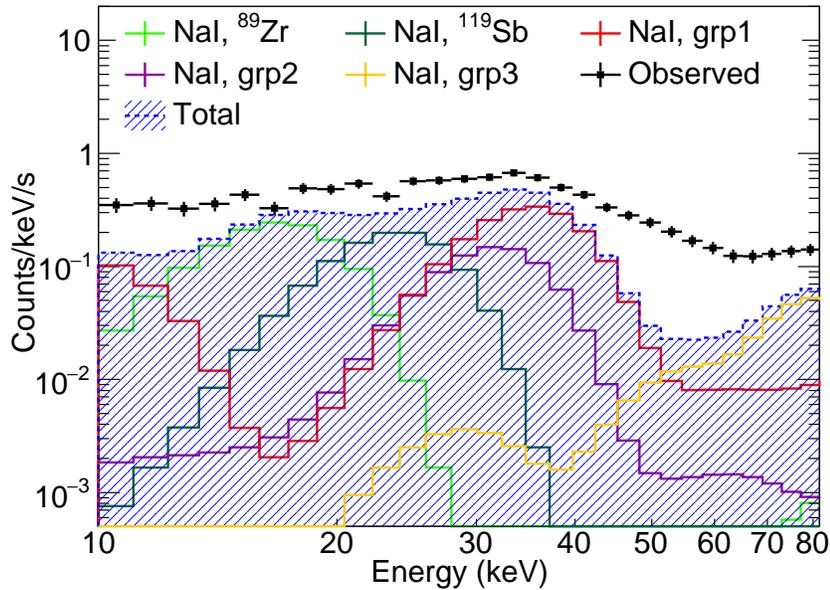}
}
\caption{Contributions from different CR induced radio-activated background
counts inside the NaI crystal and in the tantalum collimator along with the
observed detector counts at 30 km altitude. Isotope elements contributing in
grp1, grp2 and grp3 are given in Table \ref{tab:cosmoTab}.}
\label{fig:CosmoComp}
\end{figure}

The detector counts from direct energy deposition of atmospheric CR alone can
not explain the observed background. Therefore, we need to consider the
contribution to the background counts from the radionuclides induced in the
detector through Cosmogenic Radio-Activation (CRA). For this calculation, we 
considered the decay simulation of all the individual cosmogenic isotopes 
produced through the primary plus secondary GCR interactions inside the 
detector. Although, a lot of different radio-nuclides are produced inside the 
detector, all of them do not contribute to the detector counts within the 
observed energy range and observation time slot. The contribution from the CRA 
component is calculated by considering the decay of all the radionuclides 
produced in the NaI crystal as described previously in Sec. \ref{ssec:simumethod} 
and \ref{ssec:norm}, then find their relative contributions from their relative 
abundance and half lives. The normalized contribution of the cosmogenic 
radio-nuclides are calculated by fitting the total observed background with the 
simulated decay peaks, leaving only the normalization parameter to float in the 
fitting procedure as described in Sec. \ref{ssec:norm}. The major contributions 
from different cosmogenic isotopes are given in Table \ref{tab:cosmoTab}. 
Considering the detector resolution factor we rearranged the different isotopic 
contribution at similar peak energies into subgroups as shown in the table. The 
relative contributions of individual radio-nuclides within different subgroups 
are also given at the last column of Table \ref{tab:cosmoTab} (separated with 
braces). These contributions from the induced radio-nuclides in comparison with 
the observed data can be seen from Fig. \ref{fig:CosmoComp}. Here, the count 
rates and relative contributions from different (sub-grouped) peaks in Table 
\ref{tab:cosmoTab} are calculated from the normalization factors obtained from 
the fitting procedure. The contributions from different induced 
isotopes inside a subgroups given at the last column of the table are 
calculated from relative abundances of the produced isotopes and their 
corresponding decay rates depending on the half lives. Therefore, we can 
distinguish the primary contributions in each subgroup.

The major contribution to the background near 36 keV (grp1) is 
due to the pile-up energy deposition from the electron shell rearrangement 
photons following the Electron Capture (EC) process in $^{123}$I. Another 
contribution to this energy is from EC in $^{125}$I producing $^{125m}$Te [35.5
keV] which 
stabilize to the ground state by emitting a $\gamma$-ray (35.5 keV). A negligible 
amount of $^{125m}$Te [144.7 keV] also produced from direct activation which subsequently 
decay by $\gamma$ emission. $^{119}$Sb decay by emitting a $\gamma$-ray around 
25 keV (23.8 keV) following the EC process. All the isotopes in grp2 contributes to the 
background mainly by the X-ray emission due to electron shell rearrangement
after EC process. $^{89}$Zr also contributes near 17 keV by the same process.
The contribution from $^{118m}$Sb [50.8 keV] near 79 keV is due to pile-up of $\gamma$ 
emission during transition from the metastable state and X-ray emission from 
atomic shell electron transition. Finally, $^{113m}$Sn [77.3 keV] contributes near 79 keV 
by emitting a $\gamma$-ray (77.3 keV) during its transition from metastable state. 
Although, in Table \ref{tab:cosmoTab} we mention the energy value only for the 
major contribution from each activated isotopes, there may be contributions
at other energies as well, as evident from Fig. \ref{fig:CosmoComp}.

\begin{table*}
\caption{Simulated count rates (Counts/s) from different cosmogenic 
radio-activated isotopes and their relative contribution (Rel. Cont.) to the
observed counts rate. Major energy deposition peaks from radioactive decay 
process (E$_{dep}$) and half life (t$_{1/2}$) of the isotopes are also given. 
Smaller contributions have been divided in three subgroups. Relative 
contributions of each radio-nuclides in their corresponding groups is given 
in the last column separated by braces.}
\label{tab:cosmoTab}
\begin{center}
\begin{tabular*}{\textwidth}{@{\extracolsep{\fill}}lccclcl@{\hskip -6mm}l@{}}
\hline\noalign{\smallskip}
Isotopes    & A   & Z  & E$^*_{dep}$ (keV) & t$_{1/2}$ & Counts/s  &
Rel. Cont. (\%) & \\
\noalign{\smallskip}\hline\noalign{\smallskip}
$^{123}$I & 123 & 53 & 35.9   & 13.22 hour\ldelim\}{3}{0mm}[] & 
\multirow{3}{*}{[grp1] 4.33} & \multirow{3}{*}{ 19.12}\ldelim\{{3}{0mm} & 97.31 \\
$^{125}$I & 125 & 53 & 35.9   &  59.40 days    &  &  &      02.36 \\
\vspace{2mm}
$^{125m}$Te & 125 & 52 & 35.9   &  57.40 days    &  &  &     00.31  \\
$^{122}$I  & 122 & 53 & 31.2          & 03.63 min\ldelim\}{8}{0mm}[] &
\multirow{8}{*}{[grp2]  2.64} & \multirow{8}{*}{ 11.66}\ldelim\{{8}{0mm} & 38.58 \\
$^{118}$Sb  & 118 & 51 & 29.0          & 03.60 min    &        &       & 31.40 \\
$^{128}$I   & 128 & 53 & 31.2          & 24.99 min   &        &       & 15.23 \\
$^{120}$Sb  & 120 & 51 & 29.0          & 15.89 min    &        &       & 05.25 \\
$^{116}$Sb  & 116 & 51 & 29.0          & 15.80 min    &        &       & 04.43 \\
$^{112}$In  & 112 & 49 & 27.0          & 14.88 min    &        &       & 02.70 \\
$^{115}$Sb  & 115 & 51 & 29.0          & 32.10 min    &        &       & 01.63 \\
\vspace{2mm}
$^{117}$Sb  & 117 & 51 & 29.0          & 02.80 hour   &        &       & 00.59 \\
\vspace{2mm}
$^{89}$Zr  & 89  & 40 & 16.7   & 78.41 hour     & 1.95 &  8.62 &       \\
\vspace{2mm}
$^{119}$Sb & 119 & 51 & 25.0          & 38.19 hour    & 1.92 & 8.49 &       \\
$^{118m}$Sb & 118 & 51 & 79.3   & 20.6 $\mu$sec\ldelim\}{2}{0mm}[]    &
\multirow{2}{*}{[grp3] 1.14} & \multirow{2}{*}{ 5.03}\ldelim\{{2}{0mm} & 94.08 \\
\vspace{2mm}
$^{113m}$Sn & 113 & 50 & 79.3          & 21.4 min &        &       & 05.91 \\
\noalign{\smallskip}\hline
\end{tabular*}
\end{center}
\footnotesize{{$^*$ E$_{dep}$ does not refer to the actual deposited energy due 
to the decay emission, but the bin center of the detector spectral binning in 
which the energy deposition contributes.}}
\end{table*}

\subsubsection{Contribution from internal natural radioactivity}
\label{sssec:naturalCont}
The background contributions from the direct energy deposition of CR particles
and from the induced radioactivity due to high energy CR particles are not yet
able to explain the observed background completely. The missing counts near
40--50 keV region suggests other source of radiation in this region which could
be due to the radiation counts from natural radioactive elements present in
the detector crystal and other surrounding materials. The possible background
source could be from the decay of $^{210}$Pb contamination present in the
detector crystal, which decays with half life around 22 years and has a
$\gamma$-ray peak at around 46 keV, as also suggested by \cite{adhi17} for the
same type of crystal. Based on this understanding, we calculate the contribution
from $^{210}$Pb isotopes in a similar method following \cite{adhi17}. We
consider a random spatial distribution of the $^{210}$Pb isotopes inside the
whole NaI crystal and on its surface and subsequently decay them. 
We considered the summed up energy for coincident depositions from the beta 
decay of $^{210}$Pb and subsequent $\gamma$ emission from the metastable 
$^{210m}$Bi. Afterwards, we modify the recorded energy deposition by the 
detector resolution during the signal digitization. The simulation shows that 
the major contribution from the natural $^{210}$Pb decay are coming from the 
crystal surface and not from the isotope distribution inside the crystal body.

\begin{figure}[!ht]
\centering
\resizebox{\columnwidth}{!}{
\includegraphics{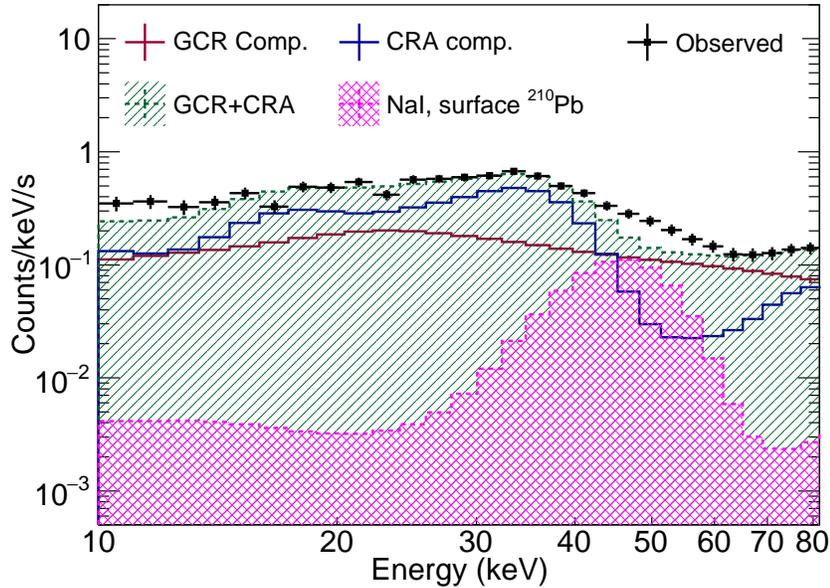}
}
\caption{Contribution of natural radioactive background counts from $^{210}$Pb
isotopes distributed on NaI crystal surface. While GCR and CRA 
components are separately shown with their combined contribution.}
\label{fig:NRComp}
\end{figure}

As we already mentioned in Sec. \ref{ssec:norm}, the normalized contribution 
of the Natural Radiogenic (NR) background component is 
obtained by fitting the overall observed background with all the components 
contributing to it (allowing the normalization parameter to float during the 
fitting). The total count rate from the decay of $^{210}$Pb 
distributed at crystal surface is 2.19 Counts/s and its relative contribution 
is 9.67\% of the total observed count rate, as obtained from the fitting. The
background contribution obtained from the simulation of natural radioactive
contamination in the NaI crystal in relation to the total observed background
counts is shown in Fig. \ref{fig:NRComp}. While separate contributions from GCR 
and CRA are also shown in the figure.

We compared the fitting results of the NR component 
with the observation data at the position of minimum counts (at time $\sim$
42000 s in Fig. \ref{fig:detLCwithAlt}a) during the balloon flight, to get an 
idea about the limit of the NR component. The analysis shows that at this 
position there are still enough counts from other sources like direct energy 
depositions from GCR and CRA component, which makes it difficult to make any 
conclusive remark about the NR component, since no absolute measurement of 
this component has not been possible on experimental basis. The NR component 
obtained from the current fitting comprises only $\sim$20.91\%
of the observed background at the position of minimum counts, compared to the
9.67\% contribution to the total background at 30 km.

\subsection{Total background contribution}
\label{ssec:totalbkg}
\begin{figure}[!ht]
\centering
\resizebox{\columnwidth}{!}{
\includegraphics{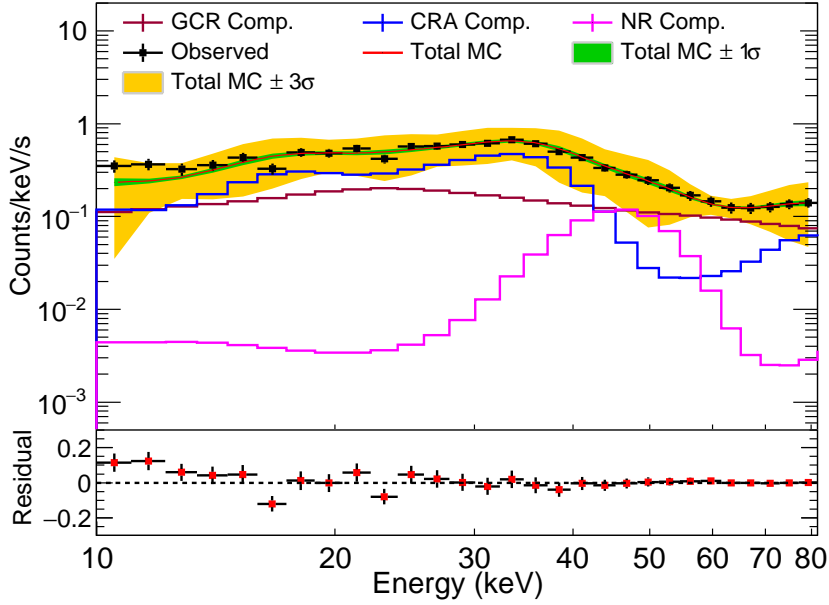}
}
\caption{The observed background counts (Observed: black points) by phoswich
detector at balloon altitude (30 km), along with the total simulated background
spectrum (Total MC: red line) and individual contributions from different
background components: Galactic Cosmic Ray component (GCR: brown histogram),
Cosmogenic Radio Activation component (CRA: blue histogram) and Natural
Radiogenic component (NR: pink histogram). The green and orange bands indicate
the $\pm$1$\sigma$ (green band) and $\pm$3$\sigma$ (orange band) statistical
uncertainty of the simulation calculation. The residual of the simulated and
observed data is shown at the bottom panel.}
\label{fig:bkg}
\end{figure}

The total background calculated from the simulation consisting all the possible
source of background fitted with the observed data is plotted in Fig.
\ref{fig:bkg}, along with the residual plot at the lower panel. All the observed
data is well within the $\pm$3$\sigma$ uncertainty of the simulation data. The
Pearson's $\chi^{2}$ test provided $\chi^{2}$/NDF = 17.33/23 with 
0.79 as the probability factor. Thus, the overall background counts 
observed in the detector is explained quite satisfactorily considering the 
components due to energy depositions from the atmospheric CR particles, 
cosmogenic radio-activation and natural radioactive isotopes in the detector. 
Here, considering the fixed (calculated) GCR component obtained from the 
simulation using realistic GCR, atmospheric and geomagnetic models, the other 
two contributors to the detector background: CRA and NR components have been 
fitted with the observed data to find their unknown contribution. The GCR, CRA 
and NR component contributes 38.73\%, 51.80\% and 9.46\% respectively to 
the total (simulated) background counts.

\section{Conclusions}
\label{sec:conc}
We have explained quite successfully the background counts from different
sources and their contributions in a scintillator X-ray detector on board a
small scientific-balloon platform. In general, the same procedure can be
followed for other X/$\gamma$-ray detectors on the ground or on board balloon or
satellites. In this procedure, we considered the primary contribution from the
GCR particles which produce other secondary particles and radiation in the
atmosphere and in turn adding to the background counts in the detector.
Albeit the detailed calculation of the radiation environment at balloon
altitude due to these particles were done by \cite{sark20}, the present
calculation shows that detector background count from direct energy depositions
of the particles and radiation is not enough to account for the whole
background. In this work, we further extend the calculation to incorporate other
effects from induced radioactivity in the detector, due to high energy CR
particles by activation and spallation of high-Z detector materials, along with
the additional contribution from the natural radioactive isotopes contaminating
the detector crystal. Since the experiment under consideration here was
performed in a solar quiet situation, the external radiation 
contribution from GCR alone (along with the cosmogenic induced and natural 
radioactivity internal to the detector crystal) could explain the detector 
background counts, without considering any contribution from sun. Otherwise, 
it may also need to consider the SEP fluxes, in particular for the detectors 
on board satellites or during high solar activity. However, 
it is worth mentioning some note of caution regarding this calculation. The 
background component from the direct GCR energy depositions is dependent on 
the models chosen to describe the GCR spectrum, atmosphere, magnetosphere etc. 
So, some systematic uncertainty in this component may persist through the 
calculation. The contribution from other two components due to CRA and NR has 
been obtained by fitting the normalization factors for both of them. Though 
the absolute contribution from CRA is very complicated to calculate due to 
diverse radiation environment and history in the atmosphere, the absolute 
contribution from the NR component, in principle, could be obtained from an 
experiment considering the detector inside a well-shielded bunker or 
inside an active shielding box (e.g., using plastic 
scintillator and PMT combination and checking for anticoincident events), 
shunning the direct GCR and CRA contributions. However, due to some 
operational constraints this experiment has not been possible yet and we rely 
on the simulation and fitting procedure to get the relative 
contributions from CRA and NR components.

\section{Acknowledgments}
The authors would like to thank the Indian Centre for Space Physics (ICSP) 
scientific balloon team members, namely, Mr. D. Bhowmick, Mr. H. Roy, 
Mr. R. C. Das and Mr. U. Sardar for their valuable supports in various forms 
during the mission operations and data collection. This work been done under 
financial support from the Science and Engineering Research Board (SERB, 
Department of Science and Technology, Government of India) project No. 
EMR/2016/003870. We also thank the Higher Education department of West Bengal, 
for a Grant-In-Aid which allowed us to carry out the research activities at 
ICSP. All the data shown in this work are available from the authors.

\bibliography{refs}

\end{document}